\documentclass[aps,prb,twocolumn,superscriptaddress,floatfix]{revtex4}
\usepackage{physics}
\usepackage{float}
\usepackage{graphicx}
\usepackage{color} 
\usepackage{amsmath} 

\begin{document}
\title{Effect of Mitigation Measures on the Spreading of COVID-19 
in Hard-Hit States}
 
\author{Ka-Ming Tam}
\affiliation{Department of Physics \& Astronomy, Louisiana State University, Baton Rouge, Louisiana 70803, USA}
\affiliation{Center for Computation \& Technology, Louisiana State University, Baton Rouge, Louisiana 70803, USA}

\author{Nicholas Walker}
\affiliation{Department of Physics \& Astronomy, Louisiana State University, Baton Rouge, Louisiana 70803, USA}

\author{Juana Moreno}
\affiliation{Department of Physics \& Astronomy, Louisiana State University, Baton Rouge, Louisiana 70803, USA}
\affiliation{Center for Computation \& Technology, Louisiana State University, Baton Rouge, Louisiana 70803, USA}

\date{\today}

\begin{abstract}

State government-mandated social distancing measures have helped to slow the growth of the COVID-19 pandemic in the United States. Current predictive models of the development of COVID-19, especially after mitigation efforts, are largely extrapolations from data collected in other countries. Since most states enacted stay-at-home orders towards the end of March, the resulting effects of social distancing should be reflected in the death and infection counts by the end of April. Using the data available through April 25\textsuperscript{th}, we investigate the change in the infection rate due to the mitigation efforts and project death and infection counts through September 2020 for some of the most heavily impacted states: New York, New Jersey, Michigan, Massachusetts, Illinois and Louisiana. We find that with the current mitigation efforts, five of those six states have reduced their base reproduction number to a value less than one, stopping the exponential growth of the pandemic. We also projected different scenarios after the mitigation is relaxed. 
Analysis for other states can be found at \url{https://covid19projection.org/}.

\end{abstract}

\maketitle

\section{Introduction}

As of April 25, there are  2.8 million confirmed cases and close to 200,000 deaths attributed to COVID-19 in the world, with over 0.9 million cases and close to 52,000 deaths in the United States. The first confirmed case of COVID-19 in the United States occurred in Washington State on January 20, 2020. Until early March, the number of reported cases remained rather low, with most of them residing in the states of Washington, New York, and California. However, since early March, the disease has spread to all of the states, with both recorded deaths and infections growing at alarming rates in many states.
Between mid-March and early-April, most states issued stay-at-home (SaH) orders.

Since the effects of social distancing measures in the United States have not be known until recently, most studies of the progress of the epidemic are largely based on extrapolations from the effects of similar strategies in other countries. However, by mid-April most states have shown signs of slowing the initial exponential growth of infection. Understanding the effects of mitigation efforts based on local data is important, as countries have implemented different degrees of social distancing measures and their effects simply cannot be translated between countries. Moreover, understanding the effect of mitigation is key to predicting the effects of relaxing those efforts. How will the timing affect the number of infections? What measures need to be enforced to keep the infection rate sufficiently low to prevent exponential growth again? For these reasons, we expand our preliminary study 
of the early stage of COVID-19 epidemic in Louisiana~\cite{Tam}. As more data is available, we now estimate the death rate and recovery rate of those in quarantine, which allows us to predict the death count, positive confirmed count, and perhaps more importantly the infected yet unidentified count after social distancing measures. 

The goal of this study is to extract the dynamics of COVID-19 in some of the most heavily impacted states and to investigate the change of the infection rate after the effects of the stay-at-home orders. We then model several scenarios with different dates for the release of the stay-at-home orders and different hypothetical increases of the infection rate. We also compare our results to the widely publicized model by the Institute for Health Metrics and Evaluation (IHME)~\cite{ihme}.

In the section II, we present the model. In the section III, we present the method of extracting the parameters for the model. In the section IV, we present the results for six stares, these include New York, New Jersey, Michigan, Illinois, Massachusetts, and Louisiana. In the section V, we discuss the error sources and the possible improvement of the projection. In the appendix, we benchmark our projection to that of the IHME. 

\section{Model} \label{Sec:Model}
We use the Susceptible-Infected-Recovered (SIR) model~\cite{Huppert,Kermack_McKendrick} , modified to consider the number of quarantined people. Similar modifications on the SIR model have been considered elsewhere to model the spread of COVID-19~\cite{Crokidakis,Bin,Pedersen,Calafiore,Bastos,Gaeta1,Gaeta2,Vrugt,Schulz,Zhang,Amaro,DellAnna,Sonnino,Notari,Simha,Acioli,Zullo,Sameni,Radulescu,Roques,Teles,Piccolomini,Brugnano,Giordano,Zlatic,Baker,Biswas,Zhang_Wang_Wang,Chen,Lloyd}.
The equations defining the dynamics of the model are as follows:

\begin{align}
\dv{S(t)}{t} &= -\beta \frac{S(t)I(t)}{N},\\
\dv{I(t)}{t} &= \beta \frac{S(t)I(t)}{N} -\qty(\alpha +\eta)I(t),\label{Eq:Ieq}\\
\dv{Q(t)}{t} &= \eta I(t)-\delta(t) Q(t) - \xi(t) Q(t),\label{Eq:Qeq}\\
\dv{R(t)}{t} &= \xi(t) Q(t) +\alpha I(t),\\
\dv{C(t)}{t} &= \delta(t) Q(t), \label{Eq:Ceq}
\end{align}

where $N$ is the total population size, $S$ is the susceptible population count, $I$ is the unidentified while infectious population count, $Q$ is the number of identified quarantined cases,  $R$ includes the number of recovered patients, and $C$ is the number of deaths.
The model is characterized by the following parameters: $\beta$ is the infection rate, $\eta$ is the detection rate, $\alpha$ is the recovery rate of asymptomatic people, $\xi$ is the recovery rate of the quarantined patients, and $\delta$ is the casualty rate of the quarantined. We further assume that all casualties had been in quarantine prior to death and we consider that only $\xi$ and $\delta$ have time dependence.



The total death count at time $t$,  $D(t)$,  can be estimated as:
\begin{equation}
    D(t) = \int_{0}^{t} \dv{C(\tau)}{\tau} \dd{\tau}.
\end{equation}
The confirmed positive count is $P(t)= Q(t)+R_Q(t)+C(t)$, where $R_Q(t)$ are the recovered patients previously in
quarantine. $P(t)$ can be estimates as:
\begin{eqnarray}
    P(t)&=& \int_{0}^{t} \dv{P(\tau)}{\tau} \dd{\tau} \\&=& \int_{0}^{t} 
    \left( \dv{Q(\tau)}{\tau} +\dv{R_Q(\tau)}{\tau}+\dv{C(\tau)}{\tau}\right) \dd{\tau}, \nonumber \\
    P(t) &=& \int_{0}^{t} \eta I(\tau) \dd{\tau}.
    \label{Eq:P(t)}
\end{eqnarray}

\section{Method} \label{Sec:Method}

We determine two sets of parameters, one before the stay-at-home order and 
the other one after the social distancing measures are in place. 
The method for estimating the model parameters from the data prior to the stay-at-home 
orders have been discussed in our previous work. \cite{Tam} We repeat our approach here for completeness of the present paper. 

Adequate testing for COVID-19 remains limited in the US. For this reason, accurately predicting the trajectory of the spread of COVID-19 by relying on the number of confirmed cases alone is a rather questionable approach, especially in the 
early stages when the percentage of people tested was very small, and the spread by infected people who were asymptomatic was significant. Alternatively, the number of fatalities attributed to COVID-19 combined with the mortality rate may be a more reliable estimator of the dynamics of the virus spread. 
Therefore, we extract the dynamics of COVID-19 from the death counts supplemented with the number of confirmed cases. 

At the beginning of the epidemic, only a small fraction of the population is infected
and we can assume the susceptible population count is very close to that of the total population, $S \sim N$. 
With this assumption one can decouple in Eq. \ref{Eq:Ieq}
the infected population count from the other variables in the model to obtain: \cite{Crokidakis,Pedersen}
\begin{equation}
\label{Eq:I(t)}
I(t) = I(0)\exp\qty[\qty(\beta -\qty(\alpha +\eta))t].
\end{equation}

At the beginning of the virus spread, the number of quarantined patients is also small
compared with the number of infected and we are able to simplify Eq.~\ref{Eq:Qeq} to 
obtain:
\begin{equation}
Q(t) = \frac{\eta}{\beta -(\alpha +\eta)}I(t). \label{Eq:QI}
\end{equation}
Combining  Eqs. \ref{Eq:QI} and \ref{Eq:Ceq}, we relate the rate of increase in the number of casualties with the number of infected people
at the early stage of the epidemic: $\displaystyle
\dv{C(t)}{t} = \delta(t) \frac{\eta}{\beta -(\alpha +\eta)} I(t) = \delta_0 I(t)$,
where $\delta_0$ is the mortality rate. Finally
the casualty count can be written as:
\begin{equation}
\label{Eq:C(t)}
C(t) = \frac{\delta_0 I(0)}{\beta-(\alpha+\eta)}\exp\qty[\qty(\beta- \qty(\alpha+ \eta))t].
\end{equation}

At the beginning of the epidemic, exponential growth of the number of fatalities is a 
reasonable assumption since the mechanisms for slowing the dynamics, such as improved detection and social distancing, are delayed in time. 
To find the initial exponent, 
$\beta-(\alpha + \eta)$,  and the prefactor, the death count and the number of deaths per day are
fit to Eq. \ref{Eq:C(t)} and its derivative $\qty(\dv{C(t)}{t})$, with
the first date with one death taken as $t=0$. We perform a
three-day moving average to smooth the data prior to the fit.  
We also discard data with less than ten deaths and use the data for the next ten days. All states in this study were still in the exponential growth phase at the last day used for the fitting of the exponent. To identify the initial number of infected people, $I(0)$, 
$\delta_0$ must be estimated.

The mortality rate, $\delta_0$, is estimated by combining the accumulated mortality rate data and the median time between infection and death. The median time between infection and the onset of symptoms is about five days, while the 
median time between the onset of symptoms and death is eight days~\cite{WHO,Anderson,Li,Linton}. It is worth noting that 
the distribution of these time intervals is close to log-normal, thus a more sophisticated analysis should include the effects of the non-self-averaging behavior of the distribution. Only the median values are used in the present work.

The accumulated mortality rate is estimated to be 2.3\% \cite{Wu}. Notably, the mortality rate does indeed vary by region. This may be due to the rate of testing as well as the capacity of health care facilities. For areas in which hospitals have been overrun, the death rate would be much higher. Notwithstanding these uncertainties, assuming that the health care facilities have not yet been overrun, the mortality rate is estimated to be $\displaystyle \delta_0 \approx \frac{0.023}{5+8} \approx 0.0018/$day.

We also estimate the recovery rate of asymptomatic people, $\alpha$, based on our current knowledge of the epidemic. Assuming that the average time to recovery or death from infection are both 13 days and that half of the infected never show any symptoms~\cite{Mizumoto}, we estimate $\alpha=0.5/13  \approx 0.0385$. This is likely closer to an upper bound of the estimate, this parameter could easily be smaller in reality. 

We estimate $\eta$ by minimizing the $\chi^2$ of the total number of deaths and confirmed cases, and their derivatives (daily number of deaths and daily number of new cases) for the last five days of the ten-day interval we are considering after the death count rise to ten deaths. After obtaining $\eta$, we can also infer the infection rate $\beta$ and the reproduction number $R_0 \approx \beta/(\eta+\alpha).$\cite{Pedersen} 

\begin{figure}[H]
\centerline{\includegraphics[width=0.5\textwidth]{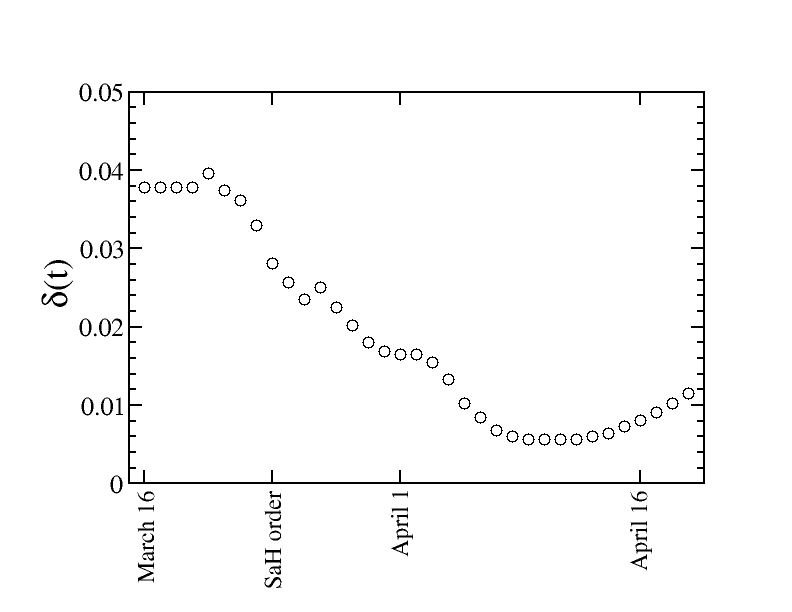}}
\caption{Estimate of the casualty rate of the quarantined cases as a 
function of time, $\delta(t)$, for the state of Louisiana.}
\label{Fig:la_delta}
\end{figure}

As opposed to our previous work \cite{Tam}, in which less data was available, we also estimate both the death rate, $\delta(t)$, 
and the recovery rate of the quarantined, $\xi(t)$, from the raw data of confirmed cases and death count as a function of time. Based on the assumption that the average time from the onset of symptoms to death or recovery is eight days, $\delta(t)+\xi(t)=1/8=0.125$. For days between the fourth and the eleventh ($t \geq 4$ and $t\leq 11$), we assume: 
\begin{equation}
\delta(t) \approx \frac{1}{8}\frac{D(t+4)-D(t-4)}{P(t-4)-D(t-4)}.
\label{Eq:delta1}
\end{equation}
For day 12 and beyond ($t\geq 12$) we assume:
\begin{equation}
\delta(t) \approx \frac{1}{8}\frac{D(t+4)-D(t-4)}{P(t-4)-P(t-12)-D(t-4)}.
\label{Eq:delta2}
\end{equation}
We assume $\delta$ for days 1 to 3 is equal to our estimate for day 4. In order to make projections we also consider that for days in the future the value of $\delta$ is equal to the value the last day with data available. This is not an unreasonable approximation, as we find the value of $\delta$ is more or less stable after the stay-at-home order becomes effective. Fig.~\ref{Fig:la_delta} displays the values of $\delta$ in Louisiana for days between March 16 and April 19. From the value of $\delta$ on April 19 we can extract a mortality rate of 2\% at this time to compare with a maximum of 6.5\% at the beginning of the epidemic. Estimates for other states behave in a similar way.

Finally, we look at the effect of the current mitigation efforts. Within the present model, there are two major routes to slow the initial
exponential growth of the epidemic,  
either to decrease the infection rate $\beta$ or to increase the testing rate $\eta$. Increasing the recovery rate of unidentified infectious people, $\alpha$, can also reduce the spread, but this is unlikely to be achieved. 
It is expected that the stay-at-home orders reduce the infection rate but do not influence the testing rate. 
However, the effect is not universal but rather highly dependent on the measures imposed. Instead of extrapolating the data from other areas, we choose to determine it from the casualty and confirmed case counts. In addition, there is a time delay for the stay-at-home order to influence the number of cases and deaths. Therefore we consider the effect of the social distancing measures to be 
reflected in two parameters, the reduction of the infection rate, $r$, and the first day when the measurements are effective, $d_{r}$. 
We determine both parameters by minimizing the $\chi^2$ of the values and daily changes of the death and the confirmed count for the five days between April 20 and 24.  

With $\delta(t)$ as calculated in Eqs. \ref{Eq:delta1} and \ref{Eq:delta2}, 
together with the initial number of infected, $I(0)$, and the rest of parameters as shown in Table \ref{Table:parameters}, we can solve the dynamics of the epidemic and estimate the death count, confirmed count, and perhaps more importantly the infected but unidentified count, $I(t)$, in each state.

\section{Results and Projections}

\begin{center}
\begin{table}
 \begin{tabular}{||r r r r r l r r r l||} 
 \hline
 State & $\beta$ & $\eta$ & $R_0$& $I(0)$ & $d_{r}$ & $r$  & $R^{SaH}_0$ &$t=0$ & SaH(t) \\  
\hline\hline
NY & 0.484 & 0.070 &4.46 & 1723 & 16(9)	& 0.13 & 0.58 & 3/15 & 3/22(7)\\
NJ & 0.436 & 0.091 &3.36 & 132  & 24(14)	& 0.20 & 0.67 & 3/11 & 3/21(10)\\
MI & 0.449 & 0.057 &4.69  & 881  & 13(8)	& 0.11 & 0.52 & 3/19 & 3/24(5)\\
MA & 0.445 & 0.130 & 2.64& 884  & 13(10)	& 0.33 & 0.87 & 3/21 & 3/24(3)\\
IL & 0.421 & 0.107 &2.89 & 481  & 14(11) & 0.35 & 1.01 & 3/18 & 3/21(3) \\
LA & 0.379 & 0.040 & 4.83 & 479  & 18(10) & 0.05 & 0.24 & 3/15 & 3/23(8) \\
 \hline
\end{tabular}
\caption{\label{Table:parameters} Parameters for different states:
the initial infection rate $\beta$, the detention 
rate $\eta$, the initial reproduction number $R_0 \approx \beta/(\eta+\alpha)$, the initial 
number of infected people at the day of the first confirmed death $I(0)$, the first date that social distancing measures are effectively stabilizing  the number of  daily deaths (in number of days since SaH order) 
$d_r$, the reduction in the infection rate $r$, the reproduction number after SaH orders
$R^{SaH}_0$, and the day of the first death and the date of the SaH order.
Massachusetts has not implemented a stay-at-home order but closed non-essential services on March 24th. The recovery rate of asymptomatic people $\alpha=0.0385$ is assumed as a constant of the model.}

\end{table}
\end{center}

\begin{figure*}
    \begin{minipage}{.495\textwidth}
        \centering
        \includegraphics[width=\textwidth]{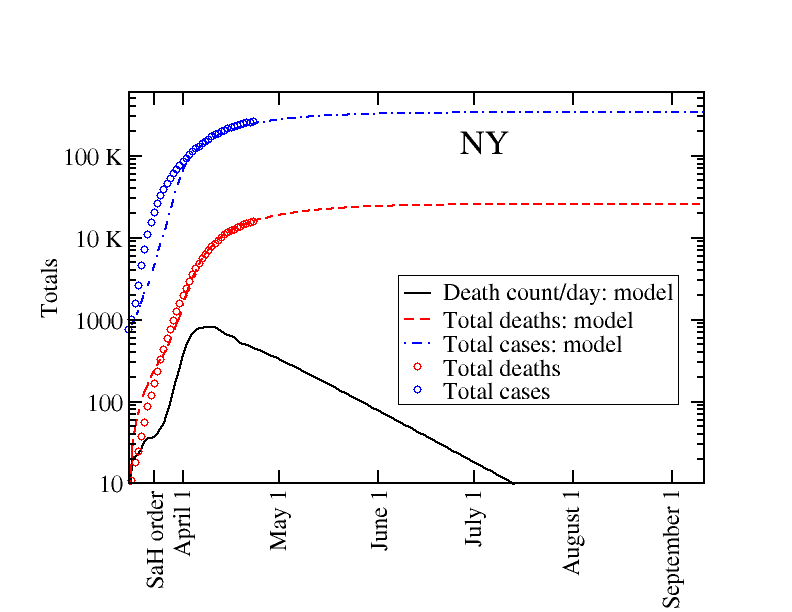}
    \end{minipage}
   \hfill
    \begin{minipage}{.495\textwidth}
        \centering
        \includegraphics[width=\textwidth]{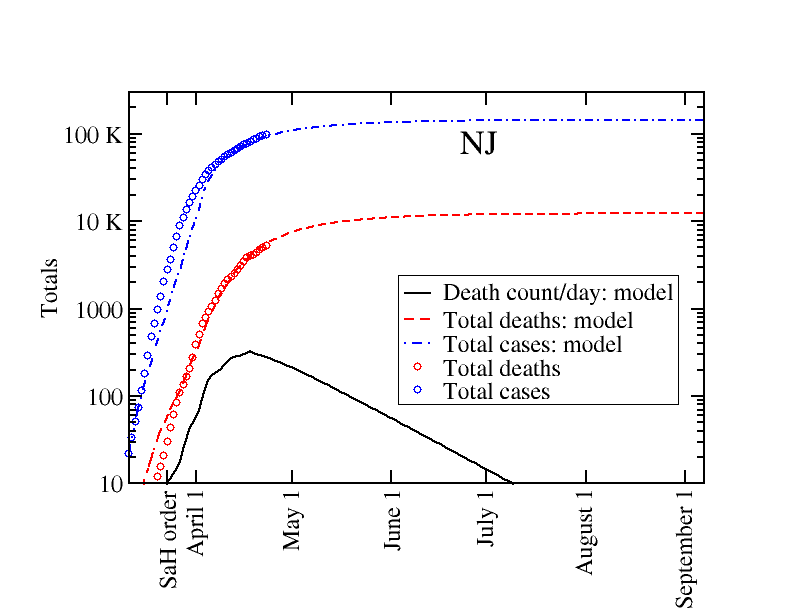}
    \end{minipage}
    \begin{minipage}{.495\textwidth}
        \centering
        \includegraphics[width=\textwidth]{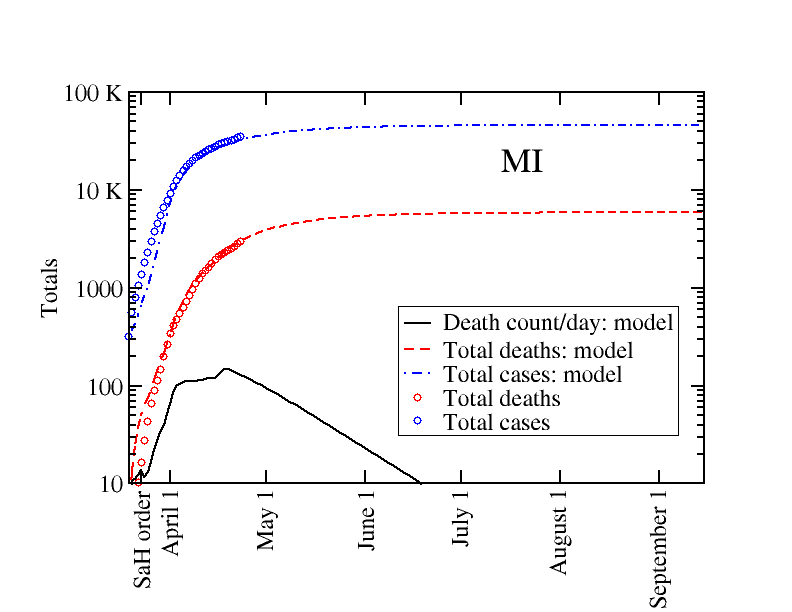}
    \end{minipage}  
       \hfill
    \begin{minipage}{.495 \textwidth}
        \centering
        \includegraphics[width=\textwidth]{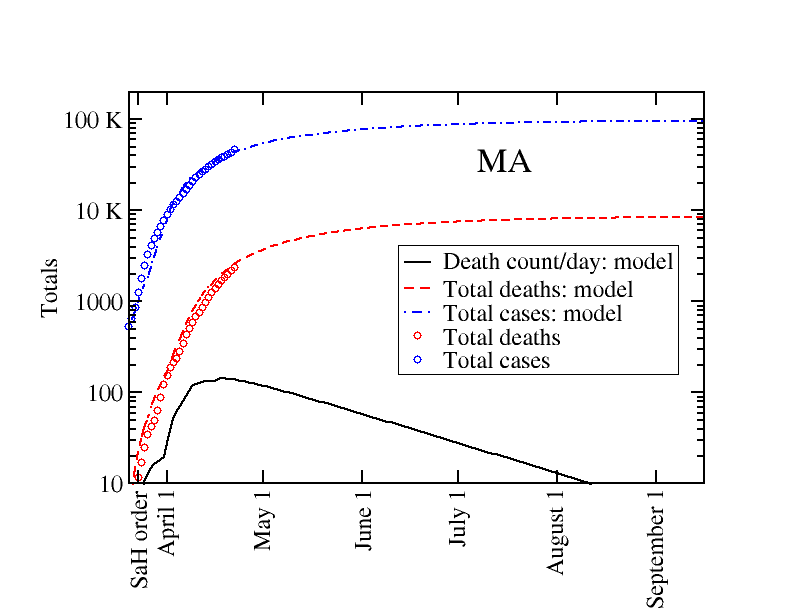}
    \end{minipage}  
    \begin{minipage}{.495\textwidth}
        \centering
        \includegraphics[width=\textwidth]{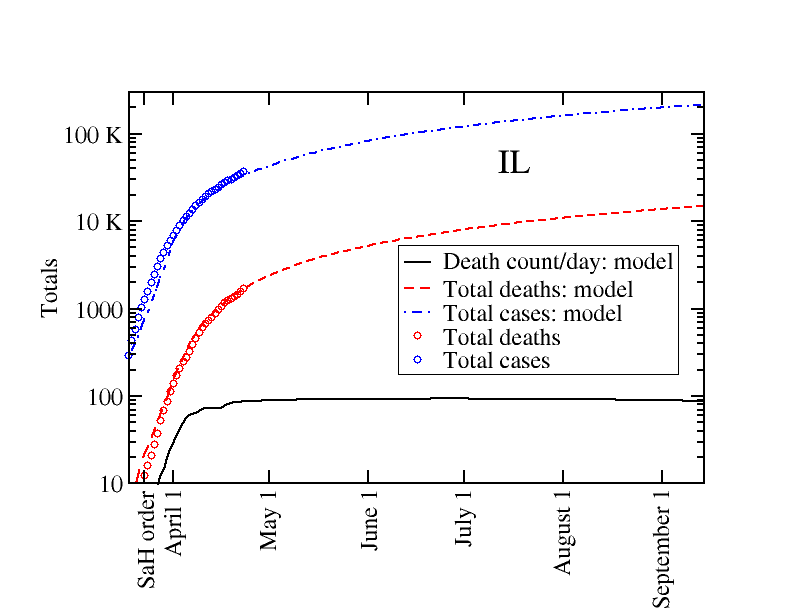}
    \end{minipage} 
         \hfill
    \begin{minipage}{.495\textwidth}
        \centering
        \includegraphics[width=\textwidth]{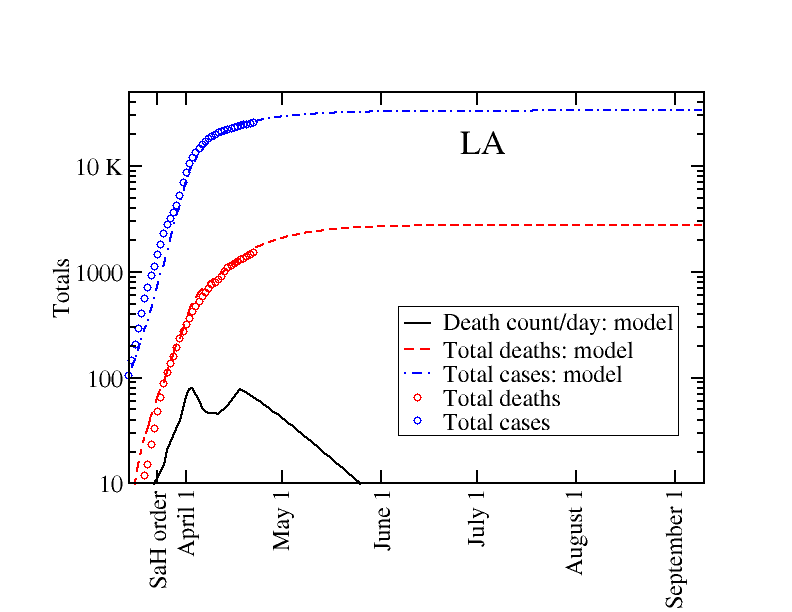}
    \end{minipage}  
    \caption{\label{fig:state-predictions} Log-scale of the daily death count (black solid curve), total number of casualties (dashed red curve),   and total number of confirmed cases (dash-dotted blue curve) 
    as function of time for six states: New York, New Jersey, Michigan, Massachusetts, 
    Illinois and Louisiana. The data for the number of deaths and cases is included as circles.} 
  
\end{figure*}

We chose six states with high death counts to test our projections. These include New York (NY), New Jersey (NJ), Michigan (MI), Massachusetts (MA), Illinois (IL), and Louisiana (LA). The casualty and the confirmed case counts are obtained from the database of the New York Times.\cite{NYT_data} Table \ref{Table:parameters} displays the
parameters of our model for these states. In particular, we can compare the reproduction 
number in the exponential growing phase of the epidemic $R_0 \approx \displaystyle \frac{\beta}{\eta+\alpha}$, with 
an effective $\displaystyle R^{SaH}_0 \approx \frac {r \beta}{\eta+\alpha}$ after the SaH 
order become effective. While the $R_0$ values are between $2.64$  for Massachusetts
and $4.83$ for Louisiana, $R^{SaH}_0$ values are  between $1.01$ for Illinois and $0.24$ for Louisiana, showing that SaH orders have been effective to reduce $R_0$ to values less than one and control the exponential growth of the disease in most states.

Fig.~\ref{fig:state-predictions} shows our predictions for the daily casualties, total number of casualties, and total number of confirmed cases for six states: New York, New Jersey, Michigan, Massachusetts,  Illinois and Louisiana. Additionally, the casualty and confirmed case counts through April 25th that are used to find the model parameters are included in the plots. First, we see that all the states have left the exponential phase and are bending their curves towards a quasi-linear region about one to two weeks after the SaH orders. However, the number of cases and fatalities in Illinois are still rapidly growing, albeit at a smaller rate than before.

The results of the model agree reasonably well with the real casualty and case counts, in particular with the casualty counts
used to estimate the initial exponent of the epidemic growth. The worst fits occurs with New York and New Jersey. Those are also the states with the largest number of cases. Difficulties in modeling New Jersey also appear in other models~\cite{marchant}. The issue might be related with the fact that NJ provides suburban housing for two large metropolitan areas, New York and Philadelphia. An analysis by metropolitan area instead of by state might be more meaningful in the case of New Jersey.
Table~\ref{Table:predictions} displays the current, as April 25, casualty and 
confirmed case counts alongside our projections through September 1. 

\begin{center}
\begin{table}[b]
 \begin{tabular}{||r r r r r ||} 
 \hline
      & Current & Current & Projected & Projected \\ 
State & (4/25)&  (04/25) &  (9/1) & (9/1) \\ 
      & deaths &  cases   & deaths & cases \\ 
      \hline \hline
NY & 16,599 & 282,174 & 25,842 & 339,826  \\ 
NJ & 5,863 & 105,523 & 12,259 & 143,672 \\ 
MI & 3,274 & 37,203 &  5,882  & 45,801   \\
MA & 2,730 & 53,348 & 8,414 & 96,000 \\ 
IL & 1,884 & 41,777 & 13,746 & 199,474  \\ 
LA & 1,644 & 26,512 & 2,789 & 33,325 \\ 
 \hline
\end{tabular}
\caption{\label{Table:predictions} The total number of casualties and confirmed cases as of April 25 and projected total deaths and cases by September 1, in six states.}
\end{table}
\end{center}

It is worthwhile to compare our results with the widely used Institute for Health Metrics and Evaluation (IHME) model~\cite{ihme}. The projected total death counts of all six states we analyze are well within the 95\% confidence interval of the IHME model on their update by Apr 25, except for Illinois. We emphasize that the present analysis is entirely based on the dynamical modeling of disease spreading with the necessary parameter inferred from death and confirmed counts alone. There is no extrapolation or interpolation of data from other countries or regions. 

\begin{figure}
\centerline{\includegraphics[width=0.5\textwidth]{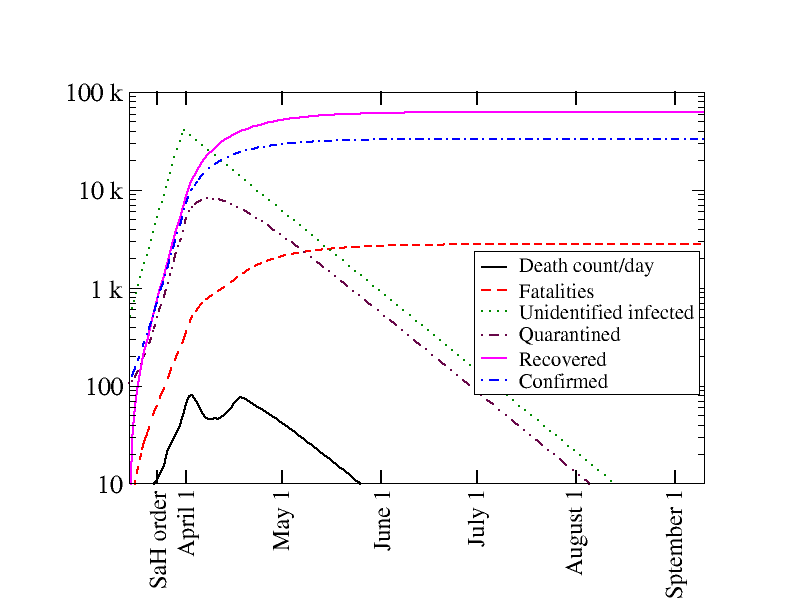}}
\caption{Model predictions for Louisiana: daily death count (solid black curve), total 
number of casualties (dashed red curve), number of unidentified infected (green dotted curve), count
of quarantined patients (double-dot-dashed maroon curve), 
total number of confirmed cases (dot-dashed blue curve), and 
total number of recovered people (solid magenta curve) as a function 
of time.} 
\label{Fig:la_full}
\end{figure}

Unlike models based on statistical inference, the present model can provide additional information on the epidemic dynamics. We focus on further analyzing the predictions for Louisiana by plotting the full set of variables. In particular, this provides a hint on the number of infected but never identified cases. Additionally, the total number of infections can also be inferred. Fig. \ref{Fig:la_full} displays the daily death count $\qty(\dv{C(t)}{t})$, total number of casualties ($D(t)$), number of unidentified infected ($I(t)$), number
of quarantined patients ($Q(t)$), total number of confirmed cases ($P(t)$), and 
total number of recovered people $\left(\displaystyle \int_{0}^{t} \dv{R(\tau)}{\tau} \dd{\tau}\right)$ 
as a function  of time. Note that by September 1, the total number of recovered people (previously in 
quarantined or unidentified) is 62,509, almost double of the 33,325 projected confirmed cases. 

A criteria to relax the social distancing measures for models based on statistical inference methods is for the number of daily confirmed case to drop below one per million  population \cite{ihme}. Since the present model consider the dynamics, we can estimate the increase in the infection and death count by proposing an increase in the infection rate due to relaxing the stay-at-home order. 

\begin{figure}
\centerline{\includegraphics[width=0.5\textwidth]{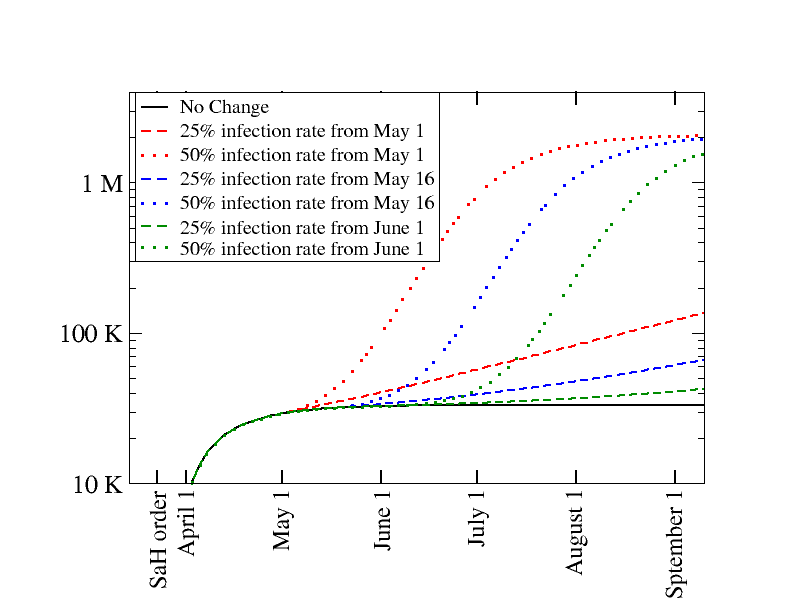}}
\caption{
 Total number of cases as a function of time for several scenarios: full mitigation efforts are in place (solid black line), the infection rate,  $\beta$, returns to 25\% (dashed curves) and 50\% (dotted curves) of its value prior to the SaH order
  at three different times, May 1, May 16, June 1.}
\label{Fig:la_open_confirmed}

\end{figure}

\begin{figure}
\centerline{\includegraphics[width=0.5\textwidth]{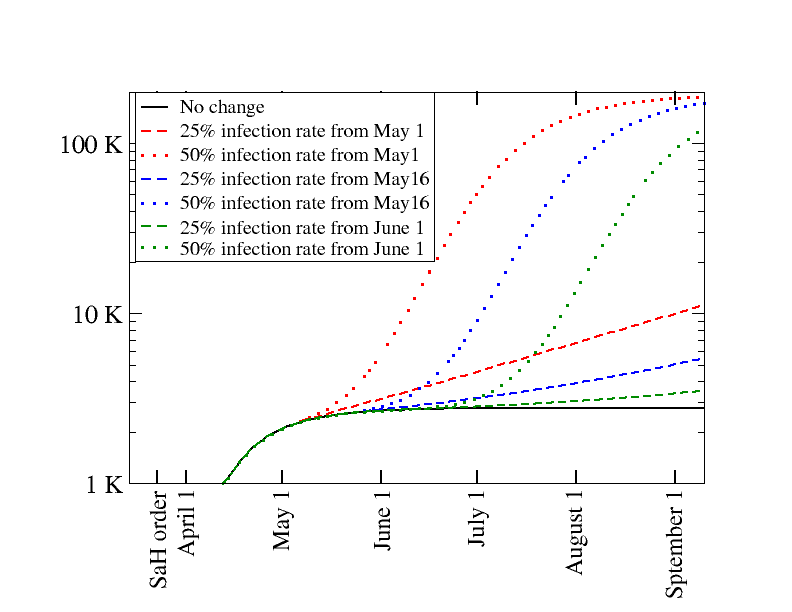}}
\caption{Total number of death counts as a function of time for several scenarios: full mitigation efforts are in place (solid black line), the infection rate,  $\beta$, returns to 25\% (dashed curves) and 50\%  (dotted curves) 
of its value prior to the stay-at-home order
  at three different times, May 1, May 16 and June 1.}
\label{Fig:la_open_death}
\end{figure}

We explore possible scenarios after the release of SaH orders for the state of Louisiana. We represent the effect of relaxing the mitigation efforts  by the increase of the infection rate, $\beta$. 
Because the number of susceptible persons, $S(t)$, has not changed sufficiently to reach herd immunity, if the value of $\beta$ reverts to the one before mitigation, the number of infected people will again grow exponentially.
We investigate the effect of increasing $\beta$ at different times, e.g. May 1, May 16, and June 1.
How much $\beta$ will increase once SaH measures are relaxed depends on various factors, such as possible limitation of mass gatherings and the proportion of the population wearing personal protective gear.
Fig. \ref{Fig:la_open_confirmed} and \ref{Fig:la_open_death} show predictions for the number of confirmed cased and fatalities, respectively, for different scenarios. We assume that the infection rate, $\beta$, increases to 25\% and 50\% of its value prior to the SaH order. We see that if $\beta$ increases to 25\%, confirmed cases and deaths grow sub-exponentially but with a larger slope than the case with full mitigation efforts. If $\beta$ increases to 50\%, both confirmed cases and fatalities will grow exponentially again. 
We notice that 
the delay on relaxing the mitigation does not help substantially to lower the number of infections in the long term.




\section{Discussion}

We analyze the  dynamics of the COVID-19 spreading on six states. 
By late April, most states have been under stay-at-home orders for almost a month, and the effects of social distancing are reflected in the data. Additionally, we are able to estimate the recovery rate and the death rate for the patients under quarantine. This allows us to expand our previous study~\cite{Tam} and 
estimate the number of infectious but unidentified people, which is largely absent in models based on statistical inference. Our results confirm the widely believed speculation that the number of infected is much larger than the confirmed positive cases. 

We find that by late April the infection rate and the effective reproduction number in all the states have been significantly reduced.
However,  Illinois is still in the phase of increasing daily death and confirmed positive counts by April 25th  (see Table \ref{Table:parameters}). 
Within this model, we do not assume that the death count is exponentially decrease
after social distancing measures, but we use the current data to 
estimate the dynamics of the disease spreading. 

We also provide possible scenarios of reopening with a focus in Louisiana. As there is no data available from the United States, 
a reasonable assumption is that the infection rate will increase after the SaH order is relaxed. 
If the infection rate returns to a value close to the one at the beginning of the epidemic, the infection will grow exponentially again. We
consider two different infection rates: 25\% and 50\% of the rate prior to the SaH order 
and three different reopening times--May 1, May 16, and June 1. Clearly, all these scenarios lead to a substantial increase of infections and deaths, but we find that the infection rate is more critical than the timing of the reopening, pointing towards the importance of effective measures to reduce the infection rate after SaH orders are lifted. Besides lowering the infection rate, the growth can also be slowed by increasing the sum of the testing rate and the recovery rate of asymptomatic persons. 
While the recovery rate is probably difficult to change, the testing can be expanded. This highlights the importance of expanding testing capacity and encouraging early testing even without severe symptoms.

There are many deficits in the present model. Improvement can be achieved by including additional factors, such as correlation with different age groups, the availability of public health care, correlation with the health condition of the population, effects of the environment such as temperature and humidity, and many others. Looking at less aggregated data, such as the data for each county or metropolitan area, might also be more meaningful than grouping the data by states which could include multiple metropolitan areas with different disease dynamics. It would be also interesting to study excess deaths instead of deaths directly produced by COVID-19.   Additionally, many of the studied quantities are not expected to be Gaussian distributed. In the present work, we take either the mean or median values of their distributions. A more sophisticated study should include the distribution of these quantities to capture  non-self-averaging effects. Similar to most approaches based on the SIR model,  we implicitly assume that the population is homogeneous and well mixed, and that infection occurs without explicit time delay. 
It is worthwhile to have a detailed comparison between the present study and methods based on statistical inference of Gaussian like distributions, such as the IHME model~\cite{ihme}.

After all, most studies of the COVID-19 spreading use highly cross-grained approximations. The detail infection mechanism at the local level is largely ignored. A truly precise approach should include the dynamics of the interactions among people at the local level. It is clear that the dynamics in  New York City cannot be the same as that at the rest of New York state. This difference while important is absent in all popular models being used for the study of the evolution of the COVID-19 pandemic. Utilizing a big data approach at the local level with graph theory should provide a more meaningful detailed analysis. Given the possibility of a second wave of infection, predictions at the local level may provide more focused mitigation approaches to minimize the economical and social impact of the pandemic.

In brief, perhaps the most timely information from this study is that reopening will definitely increase the projected number of cases and fatalities, but if the infection rate can be kept to a value much lower than the rate prior to the stay-at-home orders, the exponential growth can be avoided. Control of the infection rate seems to be a more critical factor than the timing of the reopening. We have extended our approach to all states which had more than ten COVID-19 fatalities by April 10. Interested readers can find our predictions at \url{https://covid19projection.org/}, where we will update our projections in a timely manner. In particular, early estimates
based on real data of the evolution of the spread after relaxing social distancing measure will be an essential piece of information to 
predict and control the repercussions of the pandemic.

\begin{acknowledgments}
This work is funded by the NSF EPSCoR CIMM project under award OIA-1541079. This work used the high performance computational resources provided by the Louisiana Optical Network Initiative and HPC@LSU computing. Additional support (KMT) was provided by NSF grants DMR-1728457 and OAC-1931445.
\end{acknowledgments}

\appendix*
\section{Benchmark Against The IHME Model}

Here we compare our predictions with the ones from the IHME model~\cite{ihme} for seven- and eleven-day time intervals  by subtracting from the projected death counts the total number of deaths. Since our method captures the effects of mitigation exclusively from the local data and these effects are not reflected in the data until around mid-April, both projections are for death counts on April 27. 

\begin{figure*}
    \begin{minipage}{.495\textwidth}
        \centering
        \includegraphics[width=\textwidth]{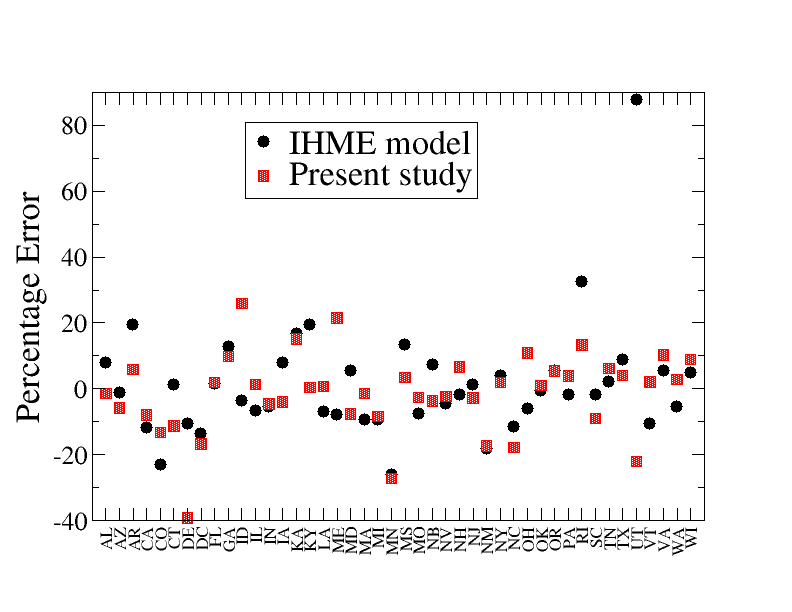}
    \end{minipage}
  \hfill
    \begin{minipage}{.495\textwidth}
        \centering
        \includegraphics[width=\textwidth]{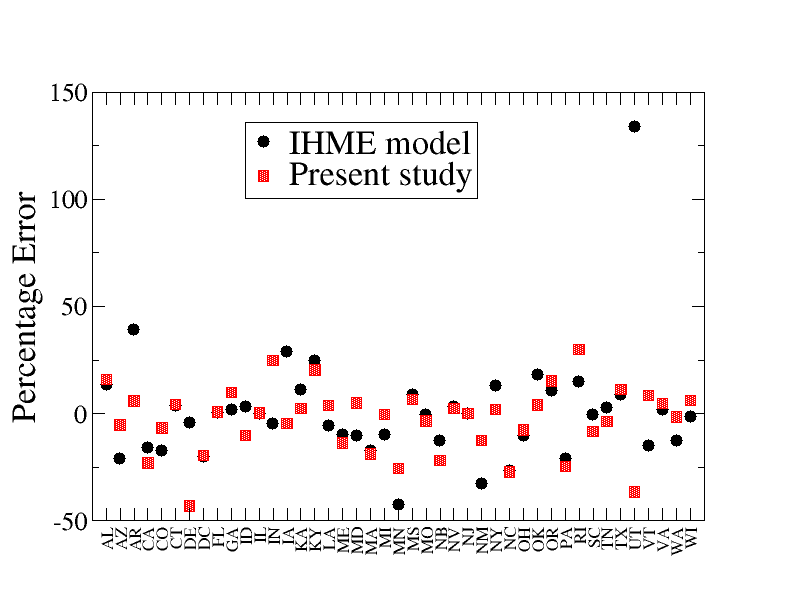}
    \end{minipage}
\caption{\label{Fig:compare}
Percentage error on the projected total death counts by April 27 of the IHME model and the present study for most U.S. states. Positive and negative values correspond to overestimates and underestimates, respectively. Note that we have subtracted 3,778 death counts from the IHME data for New York.
Right panel: seven-day projection. Data of the IHME model is from its April 21 update.  Data of  the present study is generated from data prior to the same day. 
  The average percentage error is 10.7\% for the IHME model, and 8.9\% for our approach. 
Left panel: eleven-day projection. Percentage error corresponds to the projections based on data prior to
April 17.   The average percentage error is 15.2\% for the IHME model, and 11.7\%  for our model.}   
    \end{figure*}

For the seven-day projection, we take the April 21 update from the IHME model which includes data through April 20~\cite{ihme} 
as well as our model results using the method described in Section \ref{Sec:Method} with data up to the same day. We then compare the projected total death counts of both models with the data seven days later on April 27. Given the sparsity of the data, we do not expect meaningful results can be obtained for those states which recorded less than ten total death counts by April 10 within our model and they are not considered. These states include Alaska, Hawaii, Montana, North Dakota, South Dakota, West Virginia, and Wyoming. The left panel of Fig. \ref{Fig:compare} displays the percentage error of the two models for the rest of the US states. The average percentage error is 10.7\% and 8.9\% for the IHME model and our model, respectively.

Then, we repeat the comparison for a longer term projection of eleven days. We take the April 17 update from the IHME model, which includes the data up to April 16, \cite{ihme} and our results generated with data up to the same day. We then compare the projected death counts of both models with the real data eleven days later on April 27. In this case, we also eliminate New Hampshire from the comparison since it did not record ten total deaths by April 6. The right panel of Fig. \ref{Fig:compare} shows the percentage of error for both models.  The average percentage error is 15.2\% for the IHME model, and 11.7\% for ours. We conclude that both model predictions are similar with our approach slightly outperforming the IHME model. A more thorough comparison is required to reveal the strengths and weaknesses of different models for simulating the COVID-19 spreading. 

\end{document}